\documentclass[%
preprint,
 amsmath,amssymb,
prb,
]{revtex4-2}

\usepackage{graphicx}
\usepackage{dcolumn}
\usepackage{bm}
\usepackage{hyperref}
\usepackage{siunitx}
\newcommand{\MUB}{\mu_{\text{B}}}
\usepackage{lineno}

\begin{document}


\title{Magnetic exchange interactions at the proximity of a superconductor}

\author{Uriel A. Aceves Rodriguez$^{1,2}$}
\author{Filipe Guimar\~aes${^3}$}%
\author{Sascha Brinker$^{1}$}%
\author{Samir Lounis$^{1,2}$}%
\email{s.lounis@fz-juelich.de}
\affiliation{$^{1}$Peter Gr\"unberg Institut and Institute for Advanced Simulation, Forschungszentrum J\"ulich \& JARA, 52425 J\"ulich, Germany\\
$^{2}$Faculty of Physics \& CENIDE, University of Duisburg-Essen, 47053 Duisburg, Germany\\
$^{3}$J\"{u}lich Supercomputing Centre, Forschungszentrum J\"{u}lich \& JARA, 52425 J\"{u}lich, Germany
}

\date{\today}


\begin{abstract}
Interfacing magnetism with superconductivity gives rise to a wonderful playground for intertwining key degrees of freedom: Cooper pairs, spin, charge, and spin-orbit interaction, from which emerge 
a wealth of exciting phenomena, fundamental in the nascent field of superconducting spinorbitronics and topological quantum technologies. Magnetic exchange interactions (MEI), being isotropic or chiral such as the Dzyaloshinskii-Moriya interactions (DMI), are vital in establishing the magnetic behavior at these interfaces as well as in dictating not only complex transport phenomena, but also the manifestation of topologically trivial or non-trivial objects as skyrmions, spirals, Yu-Shiba-Rusinov states and Majorana modes. Here, we propose a methodology enabling the extraction of the tensor of MEI from electronic structure simulations accounting for superconductivity. We apply our scheme to the case of a Mn layer deposited on Nb(110) surface and explore proximity-induced impact on the MEI. Tuning the superconducting order parameter, we unveil potential change of the magnetic order accompanied with chirality switching. Owing to its simple formulation, our methodology can be readily implemented in state-of-the-art frameworks capable of tackling superconductivity and magnetism. Our findings opens intriguing exploration paths, where chirality and magnetism can be engineered depending on the conducting nature of magneto-superconducting interfaces. We thus foresee implications in the simulations and prediction of topological superconducting bits as well as in cryogenic superconducting hybrid devices involving magnetic units.

\end{abstract}

\maketitle

\section{Introduction}

Despite the hostility between the superconducting and magnetic orders,  together they are known to bring to life an abundance of interesting physics, such as in-gap effects such as Majorana\cite{doi:10.1146,Kitaev_2001,Tanaka2009}, Andreev, and Yu-Shiba-Rusinov (YSR) states\cite{yu,shiba,rusinov,PhysRevB.103.205424,Tanaka2012}. These phenomena are currently in the spotlight given their potential applications in the field of topological quantum computing\cite{SciPostPhys.3.3.021,PRXQuantum.2.040347,PhysRevB.105.075129}. In the context of superconducting spinorbitronics, the interplay of the underlying Cooper pairs with the three electronic degrees of freedom---spin, charge and spin-orbit interaction---can trigger tantalizing opportunities for cryogenic quantum technologies.

Since Majorana zero modes are essential for topological quantum computing, numerous platforms have been proposed for their physical realization: magnetic islands\cite{Mnard2019}, skyrmions\cite{PhysRevB.94.064513} and spin chains\cite{Schneider2021}, among others. 
In the latter two examples, the non-collinearity of the magnetic moments is a crucial ingredient for the emergence of the coveted in-gap states. 
For non-collinearity to occur, there must be competition between the magnetic interactions in the system. 
We need thus methods to quantify and analyze these interactions within a realistic description of the electronic structure of the given systems.
Moreover, we need to understand how are these interactions affected by superconductivity and, in turn, how superconductors are influenced by the magnetic structures in their proximity.

The microscopic theory to describe conventional superconductivity goes back more than 60 years, first by the hands of John Bardeen, Leon Cooper, and Robert Schrieffer (BCS)\cite{PhysRev.106.162} in 1957. Few years later, the Bogoliubov--de Gennes (BdG) method\cite{bogol,degennes,Alaberdin1996,zhu2016bogoliubov} has been proposed and is currently an extensively used framework to investigate superconducting systems with impurities, superconductor/non-superconductor heterostructures, Josephson junctions, and topological superconductors, to name some notable examples\cite{Han_2009,PhysRevB.102.245106,Pellegrino2022,PhysRevB.105.L140504,Sato2017}. 
The BdG method is a mean-field approximation that relies upon Bogoliubov--Valatin transformations that take the Hamiltonian from a particle space into a particle-hole one, and it has been used in a variety of situations from tight-binding\cite{PhysRevB.101.134512,PhysRevB.78.024504} to density functional theory (DFT)\cite{PhysRevB.105.125143,Beck2021}. 
Especially in latter, there have been efforts to computationally analyze superconductor/non-superconductor heterostructures  based on a realistic description of the electronic structure\cite{PhysRevB.88.020407,PhysRevB.102.245106,PhysRevB.90.235433,PhysRevB.104.245415}. 
On the experimental front, the activity regarding magnetic/superconductor interfaces has been intense as well, with much focus on Majorana modes and other end-states on atomic chains\cite{Feldman2016,Schneider2020,Schneider2021,Schneider2022,doi:10.1126/sciadv.abi7291,Kster2022}. 
In this paper, we provide a simple and detailed demonstration on how to quantify the magnetic exchange interactions (MEI) from electronic structure simulations of realistic materials accounting for  electron-hole coupling channels  originating from the superconducting order in the BdG method, spin-orbit coupling and multi-orbital hybridization phenomena. We thus go beyond fundamental basic models suggested in the past\cite{oldRussianPaper,PhysRevLett.113.087202}. 
Furthermore, we apply the proposed methodology to a magnetic interface with a superconducting substrate: a Mn monolayer deposited on Nb(110) surface and analyse the different contributions to the MEI emerging from the proximity to a superconductor.

This paper is organized as follows: In Subsection~\ref{sec:bdg-theory}, we introduce our multi-orbital tight-binding theory necessary for the incorporation of the BdG equations. 
Subsection~\ref{sec:mei-theory}
contains a discussion and description of the theoretical formalism enabling the quantification of the bilinear tensor of magnetic exchanges in the context of the BdG method. 
We present a simple formula to calculate the bilinear magnetic exchanges within the Green function formalism in the particle-hole space, and establish how this formula converges to the same one as in the metallic case\cite{LIECHTENSTEIN198765,PhysRevB.79.045209}
in absence of superconductivity.
In Section~\ref{sec:calculations}, a prototypical system composed of one monolayer of Mn (110) on top of a 5-atom-thick slab of Nb (110) is used as a proof-of-concept to which we apply our theory.
We perform self-consistent calculations to investigate the effect that the superconductivity in Nb has on the magnetic properties of the Mn atoms and vice-versa. 
We find the magnetic ground state of the Mn monolayer to be row-wise antiferromagnetic, agreeing with recent experimental and theoretical results\cite{PhysRevB.105.L100406}. 
We proceed to test a wide range of electron-phonon coupling strengths that directly influences the size of the superconducting gap, evaluating the resulting self-consistent gap parameters and the corresponding magnetic moments in Mn. 
The effect of superconductivity in the Heisenberg exchange interactions for different gap sizes is scrutinized in Subsection~\ref{sec:heisenberg}. 
We find that for a realistic value of the superconducting gap, the change induced by superconductivity is minimal and therefore does not have important repercussions for the case of Mn/Nb(110).
However, for gap values of the order of the Heisenberg exchange, we observe a change in the magnetic ground state, from row-wise antiferromagnetic to ferromagnetic.
Finally, in Subsection~\ref{sec:dmi}, we focus on the Dzyaloshinskii-Moriya interaction (DMI) and observe that the corrections due to superconductivity are also relatively small in this case. 
Nevertheless, we notice that in our case the chirality of the corrective term is opposite to the one at the non-superconducting case. 
Utilizing a multiple-scattering expansion, we identify how the intertwining of the superconducting parameter, intra-atomic spin-orbit and exchange interactions impact the sign of the Heisenberg exchange as well as the DMI.

\section{Theoretical Description}\label{sec:theory}

\begin{figure}[ht!]
\centering
\includegraphics[width=\columnwidth]{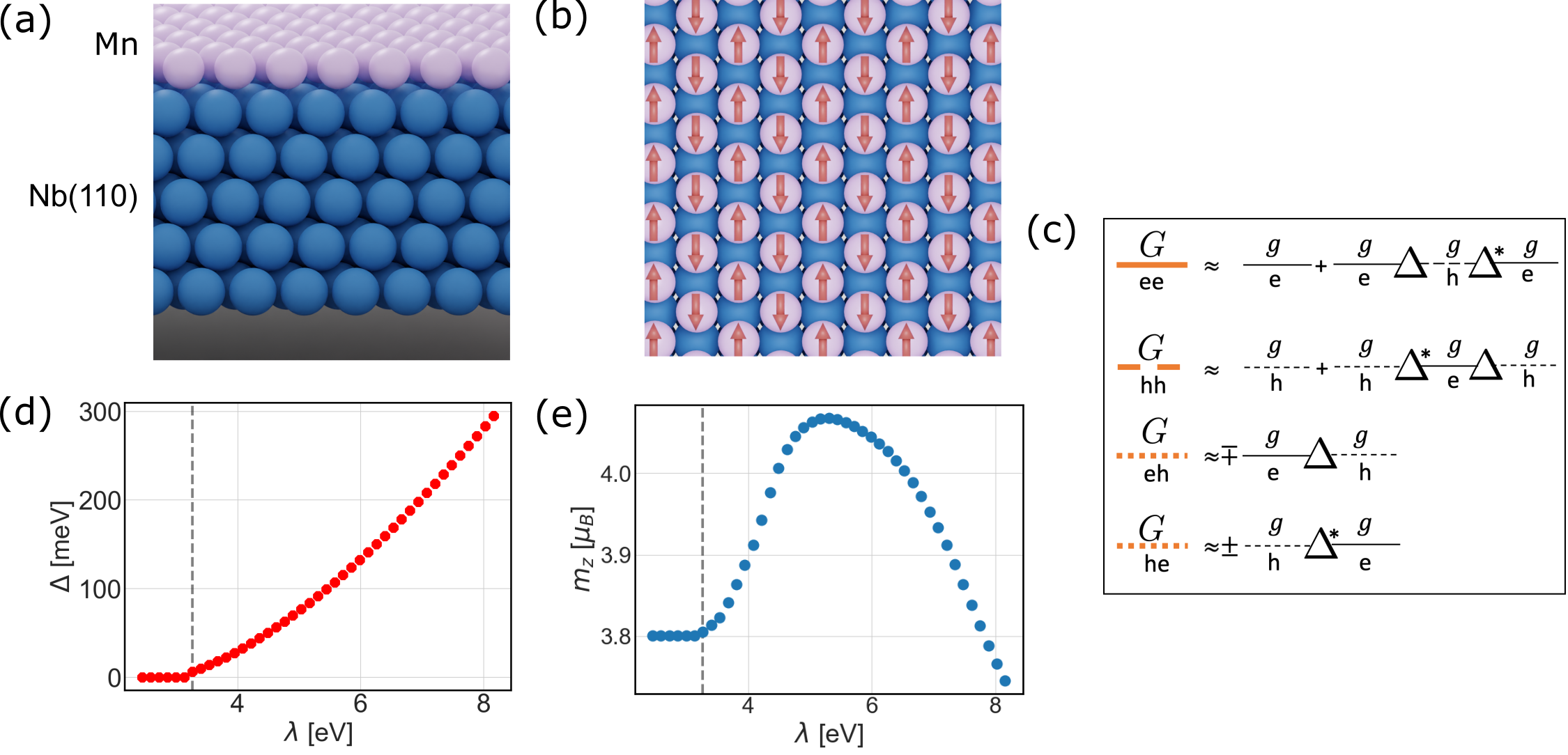}
\caption{
(a) Side view of a slab of five Nb (110) layers with a Mn monolayer on top, blue spheres represent Nb atoms, lilac spheres Mn. (b) Top view. Each layer has the structure of a centered rectangular lattice. The arrows show the magnetic ground state which for this case is row-wise antiferromagnetic. (c) Schematics of different Green functions after perturbative expansion with respect to the superconducting order parameter. (d) Superconducting gap parameter as function of $\lambda$, the vertical dotted line is at $\Delta$=\SI{3.264}{\milli\electronvolt} and indicates the lowest value of $\lambda$ for which TITAN converged to a state with a finite $\Delta$. All calculations were done at $\SI{4.2}{\kelvin}$ with 15000 k-points. (d) Magnetic moments of the Mn layer as a function of $\lambda$. 
}
\label{fig:balls}
\end{figure}

To investigate the characteristics and effects of superconducting magnetic structures, we explored a system consisting of a slab of Nb (110) with a thickness of 5 layers, with a monolayer of Mn on top, as shown in Figs.~\ref{fig:balls}(a) and (b). Fig.~\ref{fig:balls}(b) represents the magnetic ground state, which is antiferromagnetic, found after our self-consistent simulations. 
We chose Nb as a substrate given its large superconducting gap\cite{Ikushima1969} ($2\Delta=\SI{3.8}{\milli\electronvolt}$), critical temperature of $T_C = \SI{9.3}{\kelvin}$\cite{Ikushima1969}, and most importantly, given the development of recent experimental techniques to fabricate clean surfaces of Nb (110)\cite{PhysRevB.99.115437}. The latter work led to a breakthrough,  and since its publication the Nb(110) surface became a standard playground for the exploration of potential Majorana and YSR states  hosted by adatoms\cite{Kster2021,felix,Beck2021,PhysRevB.102.174504}, nanowires\cite{PhysRevB.103.235437,Schneider2021,Kster2022}, two-dimensional diluted structures~\cite{Soldini2023} and thin films\cite{doi:10.1021/acsnano.2c03965,PhysRevMaterials.6.024801,PhysRevB.105.L100406}. In the following, we describe the theoretical framework that we developed to quantify magnetic interactions at the vicinity of a superconductor.

\subsection{Tight-binding and the Bogoliubov--de Gennes method}\label{sec:bdg-theory}

Before presenting the method to extract magnetic exchange interactions, which is the focus of our work, we introduce the multi-orbital electronic Hamiltonian and the Bogoliubov--de Gennes method. 
The magnetic and superconducting system may be described by the Hamiltonian 
\begin{equation}
\begin{aligned}
    H_{S} = \frac{1}{N}\left\{\sum_{\alpha\beta,\sigma \eta, \mu \nu,\bm{k}} H_{\alpha\beta,\sigma \eta}^{\mu \nu}(\bm{k}) c^{\dagger}_{\alpha\mu\sigma}(\bm{k}) c_{\beta\nu\eta}(\bm{k}) - \sum_{\alpha,\mu\nu,\bm{k}\bm{k}'} \lambda_{\alpha\mu} c^{\dagger}_{\alpha\mu\uparrow}(\bm{k}) c^{\dagger}_{\alpha\mu\downarrow}(-\bm{k}) c_{\alpha\mu\downarrow}(-\bm{k}') c_{\alpha\mu\uparrow}(\bm{k}')\right\},
\end{aligned}
    \label{eq:bcs}
\end{equation}
with $c^{\dagger}_{\alpha\mu\sigma}(\bm{k})$ and $c_{\beta\nu\eta}(\bm{k})$ being the creation and annihilation operators of electrons with wave vector $\bm{k}$ and spin $\sigma$ in the orbitals $\mu$ or layer $\alpha$ and spin $\eta$ in the orbitals $\nu$ or layer $\beta$, respectively. $N$ is the number of wave vectors in the Brillouin zone. $\bm{k}$ is a reciprocal vector in in the xy-plane (within each layer) where two-dimensional periodic boundary conditions are assumed. 
The second element on the right-hand side corresponds to the BCS term, allowing electrons to form Cooper pairs and, therefore to give rise to superconductivity. 
Its strength, given $\lambda_{\alpha\mu}\in \mathbb{R}$, originates from the electron-phonon coupling and may depend on the orbital $\mu$ of layer $\alpha$.
$H_{\alpha\beta,\sigma \eta}^{\mu \nu}(\bm{k})$ is the non-superconducting Hamiltonian, which can be further separated into
\begin{equation}
    H^{\mu \nu}_{\alpha\beta,\sigma \eta}(\bm{k}) = H_{\alpha\beta}^{0\mu \nu}(\bm{k}) \sigma^0 + \bm{\sigma}\cdot \hat{\bm{e}}_\alpha B_\alpha^{[\text{xc}]\mu\nu}(\bm{k}) \delta_{\alpha\beta} + \bm{\sigma}\cdot \bm{B}_\alpha^{[\text{soc}]\mu\nu}(\bm{k})\delta_{\alpha\beta},
\end{equation}
where $H_{\alpha\beta}^{0\mu \nu}$ is the spin-independent tight-binding term, the second term comprises the intra-atomic exchange interactions  (originating from a Hubbard-like contribution\cite{Guimaraes2017,Guimares2020}), and the last term describe the spin-orbit interaction. The hopping parameters for Nb and Mn were obtained from first-principles calculations~\cite{papabook}. 

In the mean-field approximation, Eq.~\eqref{eq:bcs} simplifies to
\begin{equation}
\begin{aligned}
    H_{S}^{\text{MF}} = &\frac{1}{N}\sum_{\bm{k}}\left\{\sum_{\alpha\beta,\sigma \eta, \mu \nu} H_{\alpha\beta,\sigma\eta}^{\mu \nu}(\bm{k}) c^{\dagger}_{\alpha\mu\sigma}(\bm{k}) c_{\beta\nu\eta}(\bm{k})\right.\\ &- \left.\sum_{\alpha,\mu} \left(\Delta_{\alpha\mu}^{*} c_{\alpha\mu\downarrow}(-\bm{k})c_{\alpha\mu\uparrow}(\bm{k}) +  \Delta_{\alpha\mu} c_{\alpha\mu\uparrow}^\dagger(\bm{k}) c_{\alpha\mu\downarrow}^\dagger (-\bm{k}) \right)\right\},
\end{aligned}
\label{eq:mf-bcs}
\end{equation}
with
\begin{equation}
    \Delta_{\alpha\mu} = \lambda_{\alpha\mu} \frac{1}{N}\sum_{\bm{k}}\langle c_{\alpha\mu\downarrow}(-\bm{k}) c_{\alpha\mu\uparrow}(\bm{k}) \rangle, \quad \Delta_{\alpha\mu}^{*} = \lambda_{\alpha\mu} \frac{1}{N}\sum_{\bm{k}}\langle c_{\alpha\mu\uparrow}^\dagger(\bm{k}) c_{\alpha\mu\downarrow}^\dagger(-\bm{k}) \rangle,
    \label{eq:delta_lambda}
\end{equation}
$\Delta_{\alpha\mu}$ is known as the superconducting gap parameter. 
For clean superconductors it is half of the superconducting gap, as it defines the necessary energy to scatter Cooper pairs (which live at the Fermi level)\cite{tinkham2004introduction}.
It is important to note that for each choice of $\lambda_{\alpha\mu}$ and $B_i^{[\text{xc}]\mu\nu}$ the final values for $\Delta_{\alpha\mu}$ and the magnetic moments $\bm{m}^\mu$ in the ground state are obtained self-consistently.
This means that even though it seems to be linearly proportional to $\lambda_{\alpha\mu}$, this is not the case in practice as seen in Fig.~\ref{fig:balls} (d).
This leads to a more realistic characterization of materials as no determined state is enforced to the system, and they can then evolve after self-consistency into their ground state. In this work, we restrict ourselves to only two different values for $\lambda_{\alpha\mu}$, namely, a constant $\lambda_{\alpha\mu}=\lambda$ for all the orbitals of the Nb layers, and $\lambda_{\alpha\mu}=0$ for all $\mu$ in the Mn one.

To diagonalize the Hamiltonian in Eq.~\eqref{eq:mf-bcs} we use a
Bogoliubov--Valatin transformation\cite{zhu2016bogoliubov}, thus transferring the original Hamiltonian from an electron representation to an electron-hole one.  
The transformation is given by
\begin{equation}
    c_{\alpha\mu\sigma}(\bm{k}) = \sum_{n}^{'} u_{\alpha\sigma}^n(\bm{k}) \gamma_n + v_{\alpha\sigma}^{n*}(\bm{k}) \gamma_{n}^\dagger, \quad c_{\alpha\mu\sigma}^\dagger(\bm{k}) = \sum_{n}^{'} u_{\alpha\sigma}^{n*}(\bm{k}) \gamma_n^\dagger + v_{\alpha\sigma}^{n}(\bm{k}) \gamma_{n}^\dagger,
    \label{eq:bv-trans}
\end{equation}
where the tilde indicates that the sums run only over the states with positive energy\cite{zhu2016bogoliubov,de1999superconductivity}. 
This restriction in the sum is done to counteract the doubling of the degrees of freedom originated from the change of basis. 
After the transformation, we land in a system where the new particles (sometimes called \textit{bogolons}\cite{volovik2003universe}) are constituted by mixtures of electron and hole operators.
The transformation in Eq.~\eqref{eq:bv-trans} is canonical, this means that the new operators $\gamma_n$ and $\gamma_n^\dagger$ fulfill the same anticommutation relations as $c_n$ and $c_n^\dagger$, namely
\begin{equation}
    \{\gamma_n,\gamma_m\} = \{\gamma_n^\dagger,\gamma_m^\dagger\}=0, \quad \{\gamma_n^\dagger,\gamma_m\} = \delta_{nm}.
\end{equation}
After the transformation, we arrive at a set of equations of the form\cite{zhu2016bogoliubov}
\begin{equation}
    \sum_{\beta\mu} H_{\text{BdG}}^{\alpha\beta,\mu\nu}(\bm{k}) \phi_{\beta\mu}(\bm{k}) = E_n(\bm{k}) \phi_{\alpha\nu}(\bm{k}),
    \label{eq:bdg-short}
\end{equation}
where the Bogoliubov--de Gennes Hamiltonian is given by
\begin{equation}
 H_{\text{BdG}}^{\alpha\beta,\mu\nu}(\bm{k}) =\begin{pmatrix}
 H_{\alpha\beta,\uparrow\uparrow}^{\mu\nu}(\bm{k})-E_F & H_{\alpha\beta,\uparrow\downarrow}^{\mu\nu}(\bm{k})  & 0 & -\Delta_{\alpha\mu}
 \mathbb{I}\\ 
 H_{\alpha\beta,\downarrow\uparrow}^{\mu\nu}(\bm{k}) & H_{\alpha\beta,\downarrow\downarrow}^{\mu\nu}(\bm{k})-E_F  & \Delta_{\alpha\mu}
 \mathbb{I}& 0\\ 
 0 & \Delta_{\alpha\mu}^*
 \mathbb{I}& -H_{\alpha\beta,\uparrow\uparrow}^{\mu\nu *}(-\bm{k}) + E_F &-H_{\alpha\beta,\uparrow\downarrow}^{\mu\nu *}(-\bm{k})  \\ 
 -\Delta_{\alpha\mu}^*
 \mathbb{I}& 0 & -H_{\alpha\beta,\downarrow\uparrow}^{\mu\nu *} (-\bm{k}) &-H_{\alpha\beta,\downarrow\downarrow}^{\mu\nu *}(-\bm{k})  + E_F
\end{pmatrix}.
\label{eq:bdgsystem}
\end{equation}
It is important to notice that due to the transformation given in Eq.~\eqref{eq:bv-trans}, the hole-space change the wave vector arguments from $\bm{k}$ to $-\bm{k}$.
The eigenvector of Eq.~\eqref{eq:bdg-short} is
\begin{equation}
    \phi_{i\nu}(\bm{k}) = \begin{pmatrix}
u_{\alpha\nu\uparrow}(\bm{k})\\ 
u_{\alpha\nu\downarrow}(\bm{k})\\ 
v_{\alpha\nu\uparrow}(\bm{k})\\ 
v_{\alpha\nu\downarrow}(\bm{k})
\end{pmatrix}.
\end{equation}
$E_F$ in Eq.~\eqref{eq:bdgsystem} is the Fermi energy and it is placed in the diagonal such that the Fermi level is at zero for the BdG system.

Structurally we can consider $H_{\text{BdG}}^{ij,\mu\nu}$ as subdivided into four parts; namely, blocks of electron-electron, electron-hole, and hole-hole interactions. The main diagonal consists of non-hybrid interactions, while the antidiagonal terms contain only the superconducting gap parameter that hybridizes electrons and holes. For simplicity in the discussions, we break down the BdG Hamiltonian as follows
\begin{equation*}
\begin{aligned}
     H_{\alpha\beta,\mu\nu}^{\text{ee}}(\bm{k}) &=\begin{pmatrix}
 H_{\alpha\beta,\uparrow\uparrow}^{\mu\nu}(\bm{k})-E_F & H_{\alpha\beta,\uparrow\downarrow}^{\mu\nu}(\bm{k}) \\ 
 H_{\alpha\beta,\downarrow\uparrow}^{\mu\nu}(\bm{k}) & H_{\alpha\beta,\downarrow\downarrow}^{\mu\nu}(\bm{k})-E_F  
\end{pmatrix}, \quad 
&H_{\alpha\beta,\mu\nu}^{\text{eh}}=\begin{pmatrix}
 0 & -\Delta_{\alpha\mu}
 \mathbb{I}\\ 
 \Delta_{\alpha\mu}
 \mathbb{I}& 0
\end{pmatrix}, \\
H_{\alpha\beta,\mu\nu}^{\text{hh}}(\bm{k}) &=\begin{pmatrix}
  -H_{\alpha\beta,\uparrow\uparrow}^{\mu\nu *}(-\bm{k}) + E_F &-H_{\alpha\beta,\uparrow\downarrow}^{\mu\nu *}(-\bm{k})  \\ 
  -H_{\alpha\beta,\downarrow\uparrow}^{\mu\nu *}(-\bm{k})  &-H_{\alpha\beta,\downarrow\downarrow}^{\mu\nu *}(-\bm{k})  + E_F
\end{pmatrix},
&H_{\alpha\beta,\mu\nu}^{\text{he}}=\begin{pmatrix}
 0 & \Delta_{\alpha\mu}^*
 \mathbb{I}  \\ 
 -\Delta_{\alpha\mu}^*
 \mathbb{I}& 0 
\end{pmatrix}.
\end{aligned}
\end{equation*}
To avoid confusion from handling too many indices we will drop them for these submatrices whenever the context allows it. Thus, we write the BdG Hamiltonian in the following form:
\begin{equation}
    H_{\text{BdG}} = \begin{pmatrix}
    H^{\text{ee}} & H^{\text{eh}} \\
    H^{\text{he}} & H^{\text{hh}}
    \end{pmatrix}.
    \label{eq:hbdg}
\end{equation}
From Eq.~\eqref{eq:hbdg}, we can obtain the corresponding retarded Green function via
\begin{equation}
    G_{\text{BdG}}(\bm{k},E+i\eta) = (E-H_{\text{BdG}}(\bm{k}) + i\eta)^{-1}.
\end{equation}
$G_{\text{BdG}}(E+i\eta)$ in turn is a matrix that for the sake of simplicity we also consider as subdivided into four blocks as in Eq.~\eqref{eq:hbdg}
\begin{equation}
    G_{\text{BdG}} = \begin{pmatrix}
    G^{\text{ee}} & G^{\text{eh}} \\
    G^{\text{he}} & G^{\text{hh}}
    \end{pmatrix}.
    \label{eq:gbdg}
\end{equation}
When the system is not superconducting $\Delta=0$, the Hamiltonian given in Eq.~\eqref{eq:hbdg} becomes diagonal in particle-hole space (i.e., $H^{\text{eh}}$ and $H^{\text{he}}$ vanish).
Consequently $G_{\text{BdG}}$ is also diagonal ($G^{\text{eh}}=G^{\text{he}}=0$), and the non-superconducting system is described by $G^{\text{ee}}$.

\subsection{Bilinear magnetic exchange tensor and the BdG method}
\label{sec:mei-theory}

Mapping the magnetic interactions from the electronic structure simulations of a realistic system into model Hamiltonians gives us the possibility of isolating different phenomena and analysing each of them separately depending on the underlying mechanims. 
Here we focus on the extended Heisenberg model, represented by
\begin{equation}
    H_{\text{Hb}} = -\frac{1}{2} \sum_{ij} \hat{\bm{e}}_i \cdot \mathcal{J}_{ij} \cdot   \hat{\bm{e}}_j,
    \label{eq:heis}
\end{equation} 
where $ \hat{\bm{e}}_i $ is the direction of the magnetic moment for atom $i$ and $\mathcal{J}_{ij}$ is the bilinear tensor of magnetic exchange interactions with the magnetic moment for atom $j$. This Hamiltonian can be divided into three terms with different symmetries\cite{Bouaziz_2017}
\begin{equation}
    J_{ij} = \frac{\text{Tr}(\mathcal{J}_{ij})\mathbb{I}_3}{3}, \quad J_{ij}^{\text{s}} = \frac{\mathcal{J}_{ij} + \mathcal{J}_{ji} }{2} - \frac{\text{Tr}(\mathcal{J}_{ij})\mathbb{I}_3}{3}, \quad D_{ij} = \frac{\mathcal{J}_{ij} - \mathcal{J}_{ji} }{2}.
\end{equation}
Using these definitions, Eq.~\eqref{eq:heis} can also represented as
\begin{equation}
    H_{\text{Hb}}  = -\frac{1}{2}\sum_{ij}J_{ij}\hat{\bm{e}}_i \cdot \hat{\bm{e}}_j - \frac{1}{2}\sum_{ij} \hat{\bm{e}}_i \cdot J_{ij}^{\text{s}} \cdot \hat{\bm{e}}_j - \frac{1}{2}\sum_{ij} \bm{D}\cdot (\hat{\bm{e}}_i\times \hat{\bm{e}}_j).
    \label{eq:heis_extended}
\end{equation}
The first term on the right-hand side (Heisenberg exchange) favours ferro- or antiferromagnetic alignments depending on the sign of J.
The second one is the traceless anisotropic part of $\mathcal{J}_{ij}$ induced by spin-orbit coupling.
Finally, the last item in Eq.~\eqref{eq:heis_extended} corresponds to the DMI, which is finite when inversion symmetry is broken and requires spin-orbit coupling. It may induce a relative rotation in the magnetic moments of neighbouring atoms and is vital for the stabilization of magnetic textures such as skyrmions\cite{PhysRevLett.87.037203,Rler2006,Bttner2018}. 
In this work we focus only on the Heisenberg exchange and the DMI, since $J_{ij}^{\text{s}}$ is negligible.

The components of the DMI are extracted as follows\cite{Bouaziz_2017}
\begin{equation}
    D_{ij} = \begin{pmatrix}
    0 & D_z & - D_y \\
    -D_z & 0 & D_x \\
    D_y & -D_x & 0
    \end{pmatrix}.
\end{equation}
Having a realistic description of the magnetic interactions of the system through these terms is instrumental for the understanding and the development of upcoming technologies that rely on magnetism, such as spintronic devices\cite{Linder2015} with magnetic domain walls\cite{PhysRevB.90.104502}, spintronic diodes\cite{Santamaria2022,Strambini2022}, and superconductor/ferromagnet systems for quantum computing\cite{Cai2022}. 
Although the purely magnetic scenario has been intensely investigated, that is no longer the case when the structure contains a superconductor. 
Here, Cooper pairs enter the picture through the coupling between the electronic and hole states, which is not taken into account by the basic theory, developed exclusively for metals.
To analyze those cases we need to account for the potential mutual impact of superconductivity and magnetism. 

There are several techniques to obtain the tensor $\mathcal{J}_{ij}$ from electronic structure simulations, one of the most common being the infinitesimal rotations method\cite{LIECHTENSTEIN198765}, which presents a way to map energies from the electronic structure into energies on an extended Heisenberg model. 
The basic idea behind this approach is to perturb the magnetic moments at two different locations
$i$ and $j$, and quantify the resultant change in energy. To get the bilinear tensor of magnetic exchanges $\mathcal{J}_{ij}$ we take the second order term of the energy change. 
For a non-superconducting system, such change (given a perturbation potential $\delta V$) is represented by\cite{LIECHTENSTEIN198765,PhysRevB.68.104436,PhysRevB.79.045209,samir_2020}   
\begin{equation}
    \delta E = -\frac{1}{\pi} \text{Im}\int_{-\infty}^{\epsilon_F} d\epsilon \sum_{p} \frac{1}{p}\text{Tr}[G(\epsilon)\delta V]^p.
    \label{vg_metal}
\end{equation}
where $p$ describes the order of the expansion and Tr is the trace over the site, spin and orbital spaces. This formula is based on the expansion of the band energy in a perturbative fashion, which is a well established technique. In order to follow its derivation, we refer the readers to the aforementioned literature.  
The extension of the matrix space by the Bogoliubov--de Gennes equations imposes changes on the Green functions as well as on the perturbation potential. 
Within this formalism Eq.~\eqref{vg_metal} becomes
\begin{equation}
    \delta E =  \frac{1}{\pi}  \text{Im}  \int_{-\infty}^{\epsilon_F} d\epsilon \sum_p \frac{1}{p}\text{ tr}\left\{\Theta[ G_{\text{BdG}}(\epsilon)\delta V]^p\right\}.
    \label{eq:delta_e_sum}
\end{equation}
where the new trace (tr) runs over electron-hole space in addition to the site, spin and orbital ones.
$\Theta$ is a matrix in electron-hole space, whose function in Eq.~\eqref{eq:delta_e_sum} is to isolate the purely electronic terms, and it is given by
\begin{equation}
    \Theta = \begin{pmatrix} \mathbb{I} & 0 \\ 0 & 0 \end{pmatrix}
    \label{eq:theta}.
\end{equation}
The tensor of bilinear magnetic exchanges is obtained by taking the second-order term of the expansion of Eq.~\eqref{eq:delta_e_sum}, that is

\begin{equation}
\mathcal{J}_{ij}  = - \frac{\partial^2 E_{ij}}{\partial \hat{\bm{e}}_i  \partial \hat{\bm{e}}_j }    
\end{equation}
as expected from Eq.~\ref{eq:heis}, with
\begin{equation}
    \delta E_{ij} =  \frac{1}{2\pi} \Theta \text{Im} \text{Tr}  \int_{-\infty}^{\epsilon_F} d\epsilon  G_{\text{BdG}}(\epsilon) \delta V G_{\text{BdG}}(\epsilon) \delta V.
    \label{eq:de_ij}
\end{equation}
Here, the shape of the dispersion potential $\delta V$ must also be generalized by varying the magnetization orientation vector $\hat{\bm{e}}_i$ in the BdG Hamiltonian in Eq.~\eqref{eq:bdgsystem}.
This results in
\begin{equation}
    \delta V = \begin{pmatrix} \delta V^e & 0 \\ 0 & \delta V^h \end{pmatrix},
    \label{eq:dv_matrix}
\end{equation}
where 
\begin{equation}
    \delta V^h_i = -\delta V^{e*}_i = -(B^{[\text{xc}]} \bm{\sigma}\cdot \delta\bm{e}_i)^*.
    \label{eq:dv_terms}
\end{equation}
Collecting these results into Eq.~\eqref{eq:de_ij}, we obtain
\begin{equation}
    \begin{aligned}
    \mathcal{J}_{ij} = \delta E_{ij} 
  = 
-\frac{1}{2\pi} \mathrm{Im} \mathrm{Tr_{Ls}} 
  \int_{-\infty}^{\epsilon_F}&d\epsilon\, \left[
  B^\mathrm{[xc]} \bm{\sigma}\cdot\delta\bm{e}_i G^{\text{ee}}_{ij}(\epsilon)B^\mathrm{[xc]} \bm{\sigma}\cdot\delta\bm{e}_jG_{ji}^{\text{ee}}(\epsilon) \right.
  \\ &\left. \quad - B^\mathrm{[xc]} \bm{\sigma}\cdot\delta\bm{e}_i G^{\text{eh}}_{ij}(\epsilon)B^\mathrm{[xc]*} \bm{\sigma}^*\cdot\delta\bm{e}_jG_{ji}^{\text{he}}(\epsilon)\right].
    \end{aligned}
    \label{eq:jij_titan}
\end{equation}
For non-superconducting systems, $G_{ji}^{\text{he}}(\epsilon)$ and $G_{ji}^{\text{eh}}(\epsilon)$ vanish and so does the second term in the right-hand side of Eq.~\eqref{eq:jij_titan}, leading to the resulting equation that is the same as for the non-superconducting case. 
Nevertheless, it is important to notice that when $\Delta>0$, not only is the second term finite, but also the first term gets renormalized by the presence of the superconducting gap. When analyzing the resulting MEI it is often useful to separate both terms in the right hand side of Eq.~\eqref{eq:jij_titan}, here we will do so in the following form
\begin{equation}
    \begin{aligned}
    \mathcal{J}_{ij} &= \mathcal{J}^{ee}_{ij} + \mathcal{J}^{ee}_{ij},\\
    \mathcal{J}^{ee}_{ij} &= -\frac{1}{2\pi} \mathrm{Im} \mathrm{Tr_{Ls}} 
  \int_{-\infty}^{\epsilon_F}d\epsilon\,
  B^\mathrm{[xc]} \bm{\sigma}\cdot\delta\bm{e}_i G^{\text{ee}}_{ij}(\epsilon)B^\mathrm{[xc]} \bm{\sigma}\cdot\delta\bm{e}_jG_{ji}^{\text{ee}}(\epsilon),\\
  \mathcal{J}^{eh}_{ij} &= \frac{1}{2\pi} \mathrm{Im} \mathrm{Tr_{Ls}} 
  \int_{-\infty}^{\epsilon_F}d\epsilon\,
  B^\mathrm{[xc]} \bm{\sigma}\cdot\delta\bm{e}_i G^{\text{eh}}_{ij}(\epsilon)B^\mathrm{[xc]*} \bm{\sigma}^*\cdot\delta\bm{e}_jG_{ji}^{\text{he}}(\epsilon).\\
    \end{aligned}
    \label{eq:jij_separated}
\end{equation}

Utilizing perturbation theory, we expect a second order correction to $G^{\text{ee}}$ and $G^{\text{hh}}$ due to superconductivity, while the electron-hole parts of the Green function  $G^{\text{eh}}$ and $G^{\text{he}}$ would at least experience a first-order correction involving a possible sign change (see Fig.~\ref{fig:balls} (c)), which could counteract the electron-electron contribution to the magnetic exchange interaction. This can be easily grasped by utilising the perturbative expansion of the Dyson equation that related the Green function, $g$, associated to the Hamiltonian with $\Delta = 0$ to $G$ for which $\Delta$ is finite. The Dyson equation reads
\begin{equation}
    G = g + g \Delta G
\end{equation} 
thus
\begin{eqnarray}
    G^{\text{ee}} &\approx& g^{\text{ee}} + g^{\text{ee}} \Delta g^{\text{hh}} \Delta^*  g^{\text{ee}} \label{eq_Gee_delta}\\
    G^{\text{hh}} &\approx& g^{\text{hh}} + g^{\text{hh}} \Delta g^{\text{ee}} \Delta^*  g^{\text{hh}} \label{eq_Ghh_delta}\\
    G^{\text{eh}} &\approx& g^{\text{ee}} \Delta g^{\text{hh}} \label{eq_Geh_delta} \\
    G^{\text{he}} &\approx& g^{\text{hh}} \Delta^* g^{\text{ee}}\label{eq_Ghe_delta}. 
\end{eqnarray}

To summarize the theoretical scheme, we first proceed to the electronic simulations of a given material without including the superconducting order parameter. Once the ground state is found, we extract the associated Green functions and compute the tensor of magnetic exchange interactions following the infinitesimal rotation method. Knowing the interactions, we can determine the magnetic ground state. The full procedure is repeated once the superconducting order parameter is included.

\section{Results}\label{sec:calculations}

To ascertain how the superconducting gap on the material influences the magnetic states, we selected a large range of 
values for $\lambda$ on the Nb slab, ranging from $\SI{2.45}{\electronvolt}$ to $\SI{8.16}{\electronvolt}$.
The calculations were performed with a broadening of the energy levels of \SI{0.113}{\milli\electronvolt}, which should mimic a temperature of \SI{4.2}{\kelvin}.
For bulk Nb at this temperature and $\lambda=\SI{2.45}{\electronvolt}$, we found a superconducting gap parameter of $\approx\SI{1.8}{\milli\electronvolt}$, which is close to the realistic values found by experiments (from $\approx\SI{1.41}{\milli\electronvolt}$ to $\approx\SI{1.57}{\milli\electronvolt}$\cite{Novotny1975}). 

In Fig.~\ref{fig:balls} (d), we display the resulting self-consistent gap parameters $\Delta$ at the Nb layer adjacent to the Mn layer attained for each value of $\lambda$. 
For this system, we observe that the strength needed to open the superconducting gap are larger compared to the one in bulk Nb, for which $\lambda=\SI{2.45}{\electronvolt}$ already opens a gap. 
Here, the lowest electron-phonon coupling strength to produce a superconducting state is $\SI{3.26}{\electronvolt}$, resulting in a gap parameter of $\Delta=\SI{6.25}{\milli\electronvolt}$. 
The resulting growth of the self-consistent $\Delta$ with respect to $\lambda$ is not linear, and although the coupling strength is in the order of $\SI{}{\electronvolt}$, the resulting gaps are of the same order of magnitude as the ones reported experimentally\cite{PhysRevB.105.L100406}. We note that this is enabled by self-consistency of our simulations. A one-shot calculation leads to a gap of the same order than $\lambda$.

We show the immediate impact superconductivity has on the magnetic moments of the Mn atoms in Fig.~\ref{fig:balls} (e). The magnetic moment experiences a total change of about 0.25 $\mu_B$ starting from \SI{3.80}{} $\mu_B$ in the metallic regime and intriguingly increases to reach a maximum of \SI{4.07}{} $\mu_B$ when the gap is about \SI{5.30}{\milli\electronvolt} before experiencing a decrease for larger gaps. This observed behavior is induced by the non-trivial impact of $\lambda$, simulating here the electron-phonon coupling, on the electronic structure.  

\subsection{Symmetric magnetic exchange}\label{sec:heisenberg}

\begin{figure}[ht!]
\centering
\includegraphics[width=0.8\columnwidth]{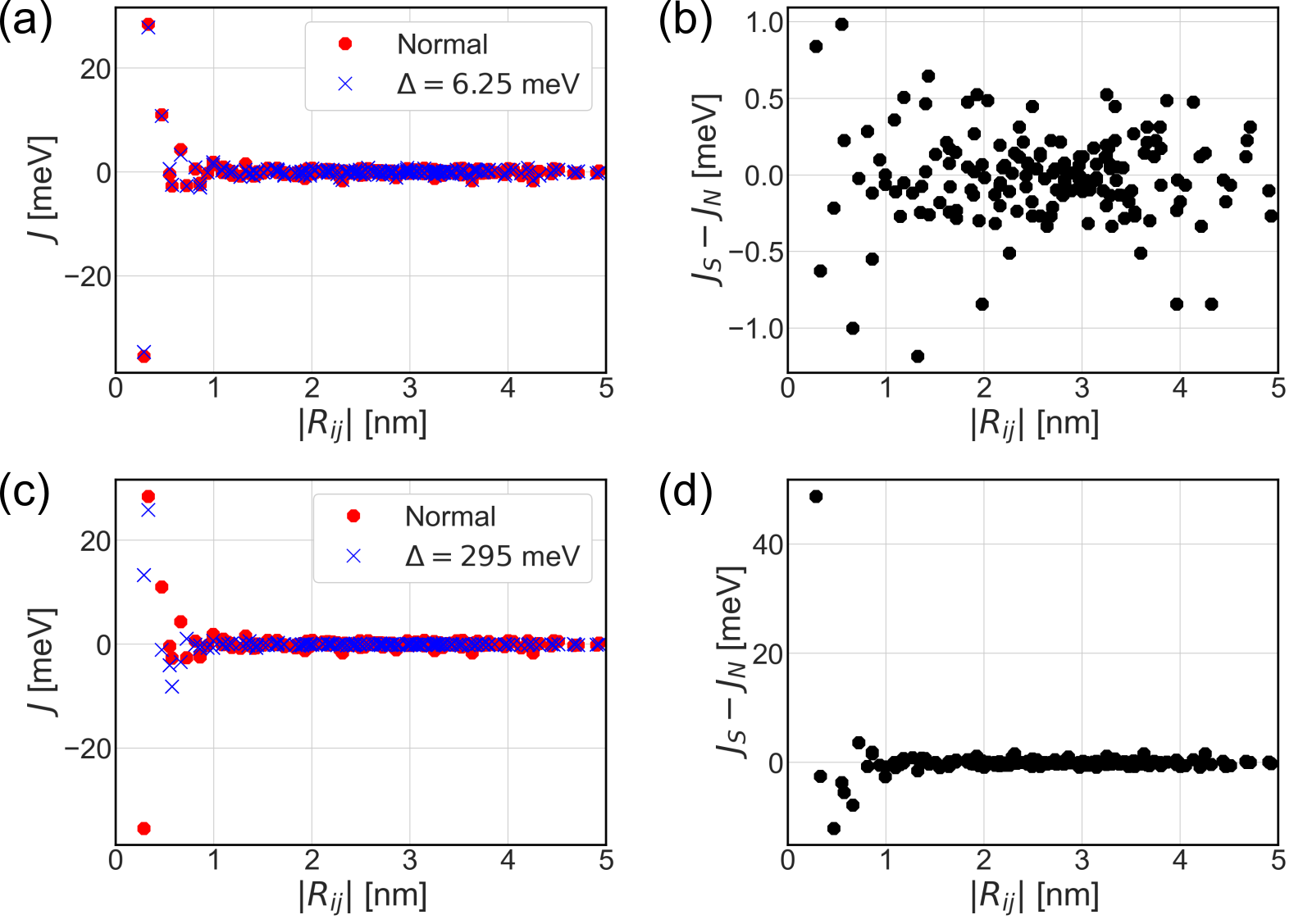}
\caption{(a) Symmetric exchange of the normal (red) vs superconducting ($\Delta=\SI{6.25}{\milli\electronvolt}$, blue) states. (b) Difference between the symmetric magnetic exchanges in the normal ($J_N$) vs the superconducting regime ($J_S$), when $\Delta=\SI{6.25}{\milli\electronvolt}$. 
(c) Symmetric exchange of the normal vs superconducting ($\Delta=\SI{295}{\milli\electronvolt}$) states. (d) Difference between $J_N$ and $J_S$ when $\Delta=\SI{295}{\milli\electronvolt}$.
All calculations were done at $\SI{4.2}{\kelvin}$, with 15,000 k-points.
}
\label{fig:j_plots}
\end{figure}

According to our convention in Eq.~\eqref{eq:heis_extended}, a positive $J_{ij}$ favours a parallel alignment as it leads to lower (negative) energies, while negative $J_{ij}$ favours anti-parallel alignment. 
The red points in Fig.~\ref{fig:j_plots} (a) represent the MEI for the normal (non-superconducting) case. While the strongest interaction coming from the nearest neighbour interaction is antiferromagnetic, the next-nearest neighbour interactions are ferromagnetic. 
Since the crystal lattice for Mn(110) is centered rectangular these MEI lead to a magnetic ground state which is row-wise antiferromagnetic as shown in Fig.~\ref{fig:balls} (b).
In Fig.~\ref{fig:j_plots} (a) we also display the Heisenberg exchange for the system when superconductivity is present (blue points). 
In this case, $\lambda=\SI{3.26}{\electronvolt}$, the smallest value in the investigated grid that leads to a finite gap parameter ($\Delta=\SI{6.25}{\milli\electronvolt}$; as a reference, the measured value for bulk Nb is in a range from $\approx\SI{1.41}{\milli\electronvolt}$ to $\SI{1.57}{\milli\electronvolt}$\cite{Novotny1975}). 

The difference between the interactions in the normal and superconducting systems is presented in Fig.~\ref{fig:j_plots} (b). 
One can notice that the maximum change in the symmetric part of the MEI is in the order of \SI{}{\milli\electronvolt}, which is small in comparison to the values of $J_{ij}$ themselves. Interestingly, enabling superconductivity at the interface does not have a uniform impact on the magnetic exchange interactions as function of distance. At short distances, a small decrease in the absolute value of the magnetic interaction is identified.
In the particularly investigated interface, however, the magnetic ground state stays unperturbed and the row-wise antiferromagnetic ordering in the Mn monolayer prevails, even after including effects derived from superconductivity. Experimentally,  Roberto lo Conte et al.\cite{PhysRevB.105.L100406} observed a row-wise antiferromagnetic ground state, where they theoretically obtain a row-wise antiferromagnetic ground state for the non-superconducting case, and experimentally observe that the magnetic ground state remains like that even when the Nb(110) slab is superconducting.

To uncover the possible effects that superconductivity may cause in the magnetic states, we artificially increase $\lambda$ to \SI{8.16}{\electronvolt} such that $\Delta=\SI{295}{\milli\electronvolt}$. Such a scenario where $\Delta$ is one order of magnitude larger than the nearest neighboring magnetic interaction looks at first sight unrealistic but it can correspond to a diluted lattice as realized recently experimentally with Cr adatoms sitting on Nb(110) surface~\cite{Soldini2023}. In that case, the magnetic interactions are rather weak since the magnetic atoms  are not nearest neighbors and interact magnetically via the substrate through RKKY-interactions. 

Although the gap is large, the magnetic moments on the Mn layer (\SI{3.75}{}$\MUB$) are still on the order of the ones in the non-superconducting case (\SI{3.80}{}$\MUB$. See Fig.~\ref{fig:balls} (e)).
The resulting Heisenberg interaction is displayed in Fig.~\ref{fig:j_plots} (c), both for the normal (red) and superconducting (blue) states. 
Their differences can be seen in Fig.~\ref{fig:j_plots} (d), in which we recognize crucial changes---the largest one being the nearest neighbour interaction that changes sign and switches from  antiferromagnetic to ferromagnetic. 
This leads to a dramatic impact on the magnetic ground state, which switches from the row-wise antiferromagnetic to a ferromagnetic one.

\begin{figure*}[bth!]
  \centering
  \includegraphics[width=1.0\columnwidth]{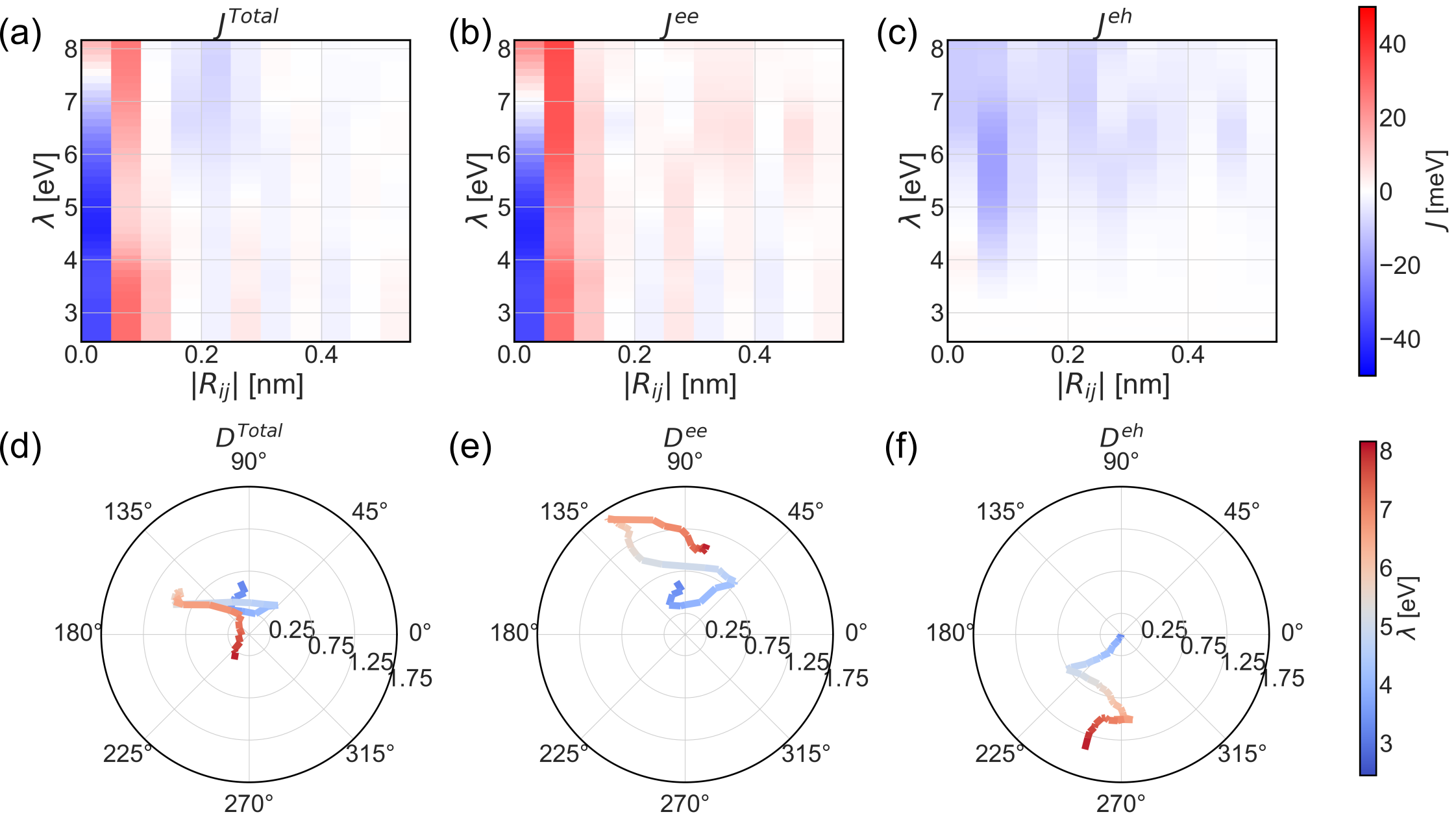}
  \caption{(a) Total exchange interaction as given by Eq.~\eqref{eq:jij_titan}, as each of its contributions separately as defined in Eq.~\eqref{eq:jij_separated}: (b) electron-electron contribution $\mathcal{J}^{ee}$ (c) electron-hole contribution $\mathcal{J}^{eh}$. (d) Polar plot ($\theta$ [Deg] vs $|\bm{D}|$ [meV]) showing the evolution of the total DMI vector coming from Eq.~\eqref{eq:jij_titan} as a function of $\lambda$ for one of the four nearest neighbours (the remaining cases are similar). (e) Evolution of the DMI vector derived from the electron-electron term $\mathcal{J}^{ee}_{ij}$. (f) Evolution of the DMI vector derived from the electron-hole term $\mathcal{J}^{eh}_{ij}$. We can see that $D^{ee}$ and $D^{eh}$ tend to have opposite orientations. All calculations were done at $\SI{4.2}{\kelvin}$, with 15,000 k-points.
  \label{fig:j_mat} }
\end{figure*}

While the previous figures show the extreme cases of low and high superconductivity in relation to the magnetic interactions, Figs.~\ref{fig:j_mat} (a)-(c) present a complete depiction of how the Heisenberg exchange as a function of the distance is affected by
the value of $\lambda$. 
The electron-hole contribution to the Heisenberg exchange for $\Delta=0$ vanishes, as expected, and hence the total Heisenberg exchange is given only by the electron-electron one. 
Additionally, we notice that the electron-hole part is mostly negative, favouring antiferromagnetism.
When focusing on the first two nearest neighbors (given by the first two columns of each plot), we detect a change in the magnetic ground state, from antiferromagnetic (blue) alignment to ferromagnetic (red). It is interesting to note that this does not originate from the electron-hole contribution, but rather from the changes experienced by the electron-electron term. These changes are caused by the shift of the bands as $\Delta$ increases.

Performing a multiple-scattering expansion of the Green functions involved in defining the magnetic exchange interactions, in the same spirit than the approach  undertaken in~\cite{Brinker2019,Brinker2020,samir_2020,Dias2021,dias2022}, we end up with systematic corrections illustrated in Fig.~\ref{fig:chiral} (a)-(b). The derivations are based on the Green functions expanded in Eqs.\ref{eq_Gee_delta,eq_Ghh_delta,eq_Geh_delta,eq_Ghe_delta}. 
For the symmetric part, we expect a second order correction due to the superconducting order parameter. Therefore the electron-electron part of the Heisenberg exchange is modified by interference effects occurring from the electron-hole and hole-electron propagation when mediating the magnetic interaction between two sites $i$ and $j$. The electron-hole part, however, inherently involves the scattering of a hole at the intra-atomic exchange of a given site $j$. such a hole-scattering is automatically accompanied with a minus sign, which could then explain the tendency of the electron-hole part to favor antiferromagnetism.

\subsection{Dzyaloshinskii–Moriya interaction}\label{sec:dmi}

Even though the possible effects of superconductivity on the Heisenberg exchange interaction may lead to interesting outcomes, one realizes soon that in magnetic systems the correction is mostly not sufficient to heavily impact the magnetic state, given that the original interaction tends to be large. 
The DMI in turn is most of the times smaller that the Heisenberg exchange, and the impact of the superconducting state might be larger there.
Although the DMI has most of the times low values compared to the Heisenberg one,
this interaction is important in the realm of non-collinear magnetism and it is responsible for stabilizing complex magnetic structures, influencing skyrmion chirality, as well as being instrumental in the study of spin waves\cite{PhysRevB.68.104436}. For the case of our system when $\Delta=\SI{6.25}{\milli\electronvolt}$ the DMI vector is small $|\bm{D}| = \SI{0.37}{\milli\electronvolt}$ with an even smaller electron-hole component $|\bm{D}^{eh}|=\SI{25.7}{\micro\electronvolt}$. Therefore, to further the impact of superconductivity we also analyzed the cases with larger superconducting gaps.
Analyzing the DMI on Mn/Nb(110), we observe that the out-of-plane component ($D_z$) is negligible, ensuing an in-plane $\bm{D}$.
\begin{figure*}[bth!]
  \centering
  \includegraphics[width=\columnwidth]{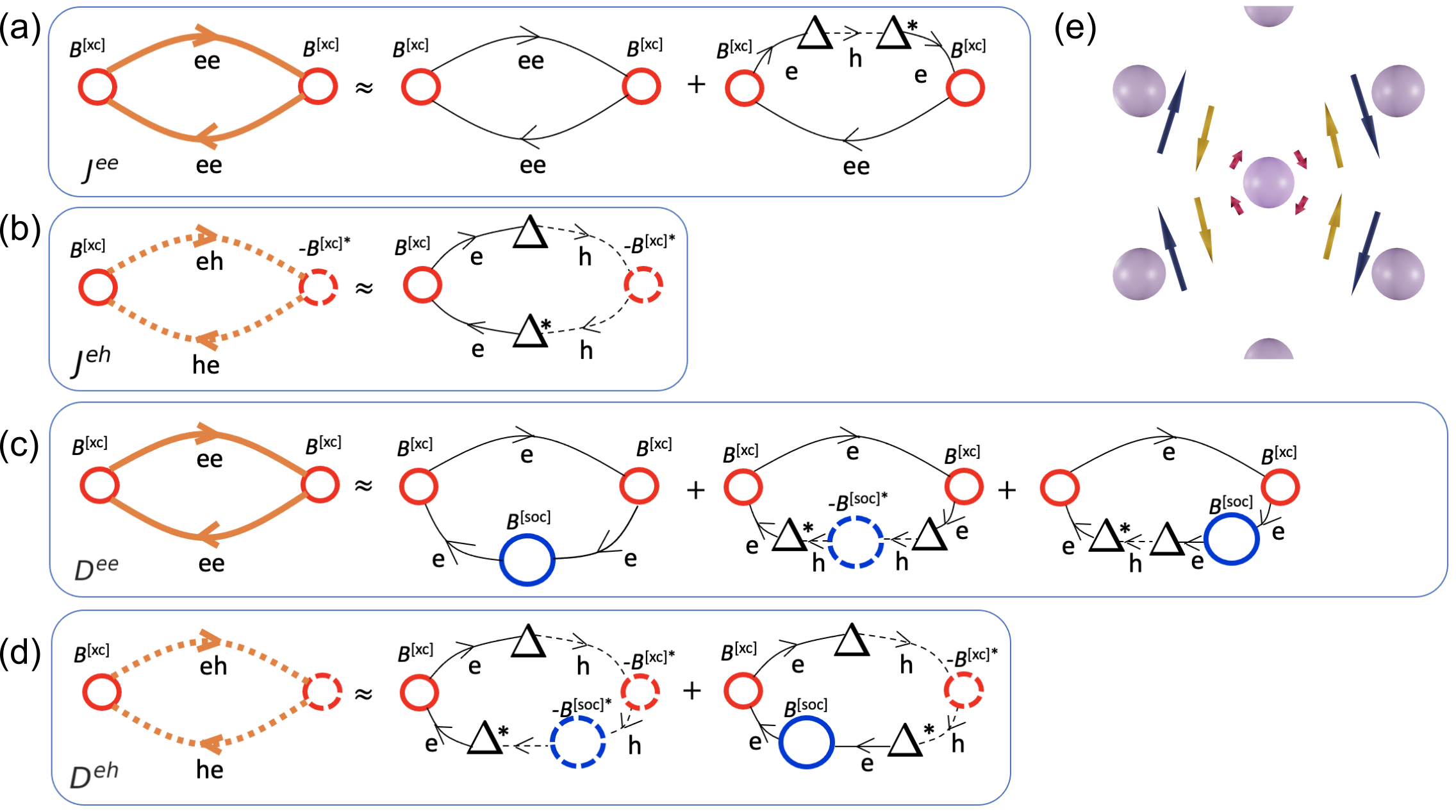}
  \caption{
  Duality of magnetic exchange interactions impacted by the superconducting order parameter. There are several scattering events that influence the exchange interactions emerging in Eqs.~\ref{eq:jij_separated}. The lines represent Green functions, thick for  finite  $\Delta$ and SOC, thin when the latter are not included. Intra-atomic exchange interactions ($B^{\text{[xc]}}$, and $-B^{\text{[xc]}}$ for electrons and holes respectively). Scattering from electron to hole ($\Delta$) and vice versa ($\Delta^*$). And finally, for the DMI there is the spin-orbit interaction ($B^{[soc]}$, and $-B^{[soc]}$ for electrons and holes respectively).
  (a) Electron-electron symmetric exchange. This term gets a second order correction with respect to $\Delta$.
  (b) Electron-hole symmetric exchange. This term has a sign change given the intra-site interaction with the hole.
  (c) Electron-electron DMI. This term has two corrections, one of them catches a minus sign coming from the spin-orbit interaction with a hole.
  (d) Electron-hole DMI. The first term gets a sign flip twice, while the second does it just once.
  (e) Total $\bm{D}$ (red), $\bm{D}^{ee}$ (golden), and $\bm{D}^{eh}$(blue) for the case with $\Delta=$\SI{295}{\milli\electronvolt}.
  \label{fig:chiral}}
\end{figure*}

In Figs.~\ref{fig:j_mat} (d)-(f), we examine the in-plane vector $\bm{D}$ for one of the four nearest neighbours in the Mn monolayer.
Here we detect that the electron-hole term tends to go in the opposite direction to the electron-electron DMI. Thus, both terms have opposite chiralities, thus, favouring different kinds of magnetic textures. 
This effect opens options such as switching between magnetic states by strengthening or weakening the superconducting order of the system at hand. 
In Fig.~\ref{fig:chiral} (e) we show $\bm{D}$ for the system with $\Delta=\SI{295}{\milli\electronvolt}$, as well as its electron-electron and electron-hole components. 
We notice that the chiralities are opposite, and the electron-hole term slightly dominates. 

Similarly to the symmetric exchange interaction, we scrutinized on the basis of perturbation theory, how the superconducting order parameter, intra-atomic exchange and spin-orbit coupling affect the DMI, as illustrated schematically in Figs.~\ref{fig:chiral} (c) and (d). The DMI emerges from the scattering at a site hosting spin-orbit coupling. Therefore the propagators in Eq.~\ref{eq:jij_separated} can be expanded in first-order with respect to spin-orbit coupling, which leads to the different processes depicted in Figs.~\ref{fig:chiral} (c) and (d). 
We notice that besides possible interference effects induced by the different electron and hole propagators, the hole-scattering at an intra-atomic exchange interaction can provide a sign change to the original chirality pertaining to the electron-electron DMI, while this can happen for the electron-hole part due to  hole-scattering at the intra-atomic exchange and spin-orbit coupling.

\section{Conclusions}

We have introduced a method for extracting the bilinear tensor of magnetic exchange interactions within the Bogoliubov-de Gennes (BdG) formalism, utilizing infinitesimal rotation of magnetic moments. This novel approach has provided insights into the intricate interplay between superconductivity, magnetism, and spin-orbit interaction, unraveling remarkable potential effects on both the Heisenberg exchange and the Dzyaloshinskii-Moriya interactions.

Through rigorous self-consistent simulations based on parameters derived from ab initio calculations, our investigation has captured the intricacies of the electronic structure in the Mn monolayer on Nb(110) surface system. By tuning the electron-phonon coupling and thereby the superconducting state, we have demonstrated the potential pivotal influence of the superconducting state on the magnetic ground state, specifically through the Heisenberg exchange. Furthermore, intriguing modifications in the chirality of the Dzyaloshinskii-Moriya interactions have been unveiled. Within the confines of Mn monolayer deposited on Nb(110) and for experimentally consistent gap sizes, we have concluded that the impact of superconductivity on the magnetic ground state remains minimal, leaving it largely unperturbed.
However, the implications of our findings could be of importance for the case where the superconducting gap is of the order of magnitude than the magnetic exchange interactions. This could be realized in the context of diluted magnetic structures, where the magnetic atoms are not nearest neighbors, as realized recently with Cr adatoms on Nb(110) surface~\cite{Kster2022,Soldini2023}.

The versatility of our method enables its application in various schemes based on Green function techniques, thereby facilitating exploration of complex magneto-superconducting interfaces. These endeavors hold immense potential for uncovering novel and non-trivial effects of superconductivity in topological magnetic and superconducting materials, where intricate magnetic interactions may play a decisive role. The exploration of complex magnetic structures, such as skyrmions or spin chains, in conjunction with superconductivity, opens up possibilities for the discovery of novel states and quasiparticles with implications for topological quantum computing,  driving the progress of cutting-edge technological applications.

\begin{acknowledgments}
The authors acknowledge  funding provided by the Priority Programmes SPP 2137 “Skyrmionics” (Projects LO 1659/8-1) of the Deutsche Forschungsgemeinschaft (DFG). 
We acknowledge the computing time provided through JARA on the supercomputers JURECA\cite{Krause_2018}. Simulations were also performed with computing resources granted by RWTH Aachen University under project jara0189.
\end{acknowledgments}

\bibliography{bibliography}

\end{document}